\documentclass[aps,prl,twocolumn,showpacs]{revtex4}

\usepackage{amsmath}
\usepackage{subeqnarray}
\usepackage{graphicx}	
\usepackage{epstopdf}

\begin{document}


\title{Core-\textit{crust} transition pressure evolution in \textit{post-glitch} epoch for a Vela-type pulsar}

\author{L. M. Gonz\'alez-Romero}
\author{J. L. Bl\'azquez-Salcedo}
\affiliation{
Depto. F\'{\i}sica Te\'orica II, Facultad de  Ciencias F\' \i sicas,\\
Universidad Complutense de Madrid, 28040-Madrid}

\date{\today}

\begin{abstract}
We propose that the \textit{post-glitch} epoch in a Vela-type pulsar corresponds to a transition time in which the broken \textit{crust} is readjusting its temperature and core-\textit{crust}  transition pressure  after the changes produced by the \textit{glitch}. To describe the evolution of the pulsar in this epoch we use a sequence of stationary and axisymmetric relativistic models of slowly rotating neutron stars with a surface layer \textit{crust}. Previously, we have formulated the matching conditions on the surface of the star in terms of  physical properties (total mass of the star, core-\textit{crust} transition pressure,...).
\end{abstract}


\maketitle



Many pulsars show sudden spin jumps, \textit{glitches}, which are superimposed to the gradual spin down due to the continued loss of angular momentum suffered by the star \cite{exp} . After a relaxation period of the order of hundreds of days (\textit{post-glitch} epoch), the star almost recovers its usual rhythm of spin down. The study of the properties of both \textit{glitch} and \textit{post-glitch} epoch are essential to understand the structure of the neutron star that models the pulsar \cite{model1}.

\textit{Glitches} have been observed in most of the younger isolated pulsars. Particularly, more than a dozen \textit{glitches} with period variation of the order of $10^{-8}$ and $10^{-6}$ have been observed in the Crab and Vela pulsars respectively since their discovery. Many mechanisms have been proposed to explain \textit{glitches} \cite{model_glitch} and \textit{post-glitch} epoch \cite{post_glitch}.

Most of the models that deal with \textit{glitches} are associated with the layer structure that neutron stars are thought to posses. Basically two regions of the star interior can be differentiated: the core and the \textit{crust}. The properties of the \textit{crust} are very different to those of the core, and it is thought that this region has a solid crystalline structure similar to a metal \cite{Haensel:2004nu}.  

Let us mention that in the thermal \textit{glitch} model proposed by Link and Epstein \cite{model_glitch}  the \textit{glitch} is explained by a sudden deposition of thermal energy in the inner \textit{crust}, energy that would come from a rupture of the solid \textit{crust}. The inner \textit{crust}  superfluid is very sensitive to star quake generated temperature perturbations, affecting to the rotation behavior of the star. Furthermore, it is noteworthy that   the dynamical properties of the neutron star depend strongly on the transition pressure  between the star's core and the \textit{crust} as have been pointed up by  Cheng, Yuan, and  Zhang \cite{Cheng} and Lattimer  and Prakash \cite{Lattimer}.

According with those two ideas, we propose  that the \textit{post-glitch} epoch can be understood as a transition time in which the broken \textit{crust} is readjusting its temperature. The temperature excess that the inner layers of the \textit{crust} has just after a  \textit{glitch}, would gradually dissipate during the \textit{post-glitch} epoch until the \textit{crust} reaches its equilibrium temperature. Our hypothesis is that the cooling of the exterior layers of the star would lead to a sensitive variation of the transition pressure between the core and the \textit{crust}.  In consequence, the \textit{post-glitch} epoch will correspond to a readjustment epoch for this transition pressure. Because many dynamical properties of the neutron star  strongly depend on the core-\textit{crust} transition pressure, the gradual cooling down of the inner layers of the \textit{crust} would explain its evolution  all along the \textit{post-glitch} epoch.

In order to be able to control the core-\textit{crust} transition pressure, we
will consider that the \textit{crust} is thin enough so that we can treat it
as a surface layer that envelops the star's core. We will consider the case of a relativistic star with rigid and slow rotation, so we can make use of Hartle and Thorne formalism \cite{Hartle}.  The introduction of  a surface layer of energy  density will change the boundary conditions on the surface of the star. We will show that it is interesting to use   the total mass, the central density and the core-\textit{crust} transition pressure as the parameters that  characterize a configuration.

 When performing the simulation of the \textit{post-glitch} epoch, we must take into account that there are two very different time scales in the evolution of the star: the \textit{post-glitch} epoch duration, that is of the order of hundreds of days, in contrast with the rotation period of the star that is of the order of seconds or even milliseconds. This important feature allows us to approximate each evolution state of the star during the \textit{post-glitch} epoch, using an hydrodynamic equilibrium configuration with the appropriate parameters (total mass, central density, and core-\textit{crust} transition pressure). We will fit the evolution of the angular velocity of the star with a sequence of configurations of different core-\textit{crust} transition pressure, assuming that neither the mass nor the central density changes during the \textit{post-glitch} epoch.
 
\begin{center}
\textbf{Model of a slowly rotating neutron star with a surface layer \textit{crust}} 
\end{center}

We assume that the neutron star is in permanent rigid rotation. The star rotates with constant angular velocity $\Omega$ around an axis, so the resulting space-time is axisymmetric and stationary. Appropriate coordinates can be chosen so that the metric can be written as follows $ds^{2}=-H^{2}dt^{2}+Q^{2}dr^{2}+r^{2}K^{2}\left[ d\theta ^{2}-\sin ^{2}\theta(d\phi -Ldt)^{2}\right].$
 We assume that the rotation of the star is sufficiently slow and, hence, the Hartle-Thorne perturbative solution can be used \cite{Hartle}. To second order in the angular velocity of the star $\Omega$, it can be proved that the components of the metric have the following expression:
\begin{subeqnarray}
g_{tt} &=&-e^{2\psi (r)}[1+2h_{0}(r)+2h_{2}(r)P_{2}(\theta )]\\&+&r^{2}\sin
^{2}\theta \omega ^{2}(r) \\
g_{t\phi } &=&g_{\phi t}=-r^{2}\sin ^{2}(\theta )\omega (r) \\
g_{rr} &=&e^{2\lambda (r)}[1+2m_{0}(r)+2m_{2}(r)P_{2}(\theta )] \\
g_{\theta \theta } &=&r^{2}[1+2k_{2}(r)P_{2}(\theta )] \\
g_{\phi \phi } &=&\sin ^{2}(\theta )g_{\theta \theta }
\end{subeqnarray}
We have to solve the Einstein equations in the interior of the star, where the matter is described by a perfect fluid tensor 
$T_{\mu \nu}=(\rho + p) u_\mu u_\nu + p g_{\mu\nu}$  with equation of state $\rho=\rho(p)$ (here $u^\mu$ is the four-velocity, $\rho$ is the proper energy density, and $p$ is the pressure of the fluid)  and in the outer region, where space is empty.  Also, we have to impose appropriate boundary conditions in the surface of the star. 

It is very desirable to introduce a new radial coordinate $R$. We can define a coordinate transformation $r \rightarrow r(R,\theta )=R+\xi (R,\theta )$ \cite{Hartle}, so that surfaces of $R=const.$ has constant density, i.e., this coordinate is adapted to the surface of the star.

We assume that the core of the star is surrounded by a thin surface layer (simplified \textit{crust} model). Then we find that in the surface of the star $R=a$, the stress-energy tensor is non null and it can be written as follows:
\begin{equation}
T_{c}^{\mu \nu }(R)=\rho _{c}(\theta ) u_{c}^{\mu }u_{c}^{\nu }(a)\delta (R-a)
\end{equation}
Hence,  we also assume that the matter of the surface layer can be described as a perfect fluid. We allow this fluid to rotate with its own angular velocity $\Omega_{c}$. For the surface energy densities we will consider, we can despise the contribution of the pressure to the surface stress-energy tensor. Therefore the surface energy density can be written to second order as follows:
\begin{equation}
\rho _{c}(\theta )=\varepsilon +\delta \varepsilon (\theta )=\varepsilon
+\delta \varepsilon _{0}+\delta \varepsilon _{2}P_{2}(\theta )
\end{equation}
where $\varepsilon,  \delta \varepsilon _{0},$ and $\delta \varepsilon _{2}$ are constants.

We must impose certain junction conditions on the interior and exterior solutions taking into account the introduction of a surface layer of matter on the border of the star. We use and intrinsic formulation of these conditions:  the first fundamental form of the hypersurface $R=a$ has to be continuous and    the second fundamental form has to have a jump given by the stress-energy tensor on the surface of the star \cite{pegado}. Expanding these junction conditions in terms of $\Omega$, we find the conditions for each one of the metric functions

\begin{figure}
\includegraphics[width=0.35\textwidth]{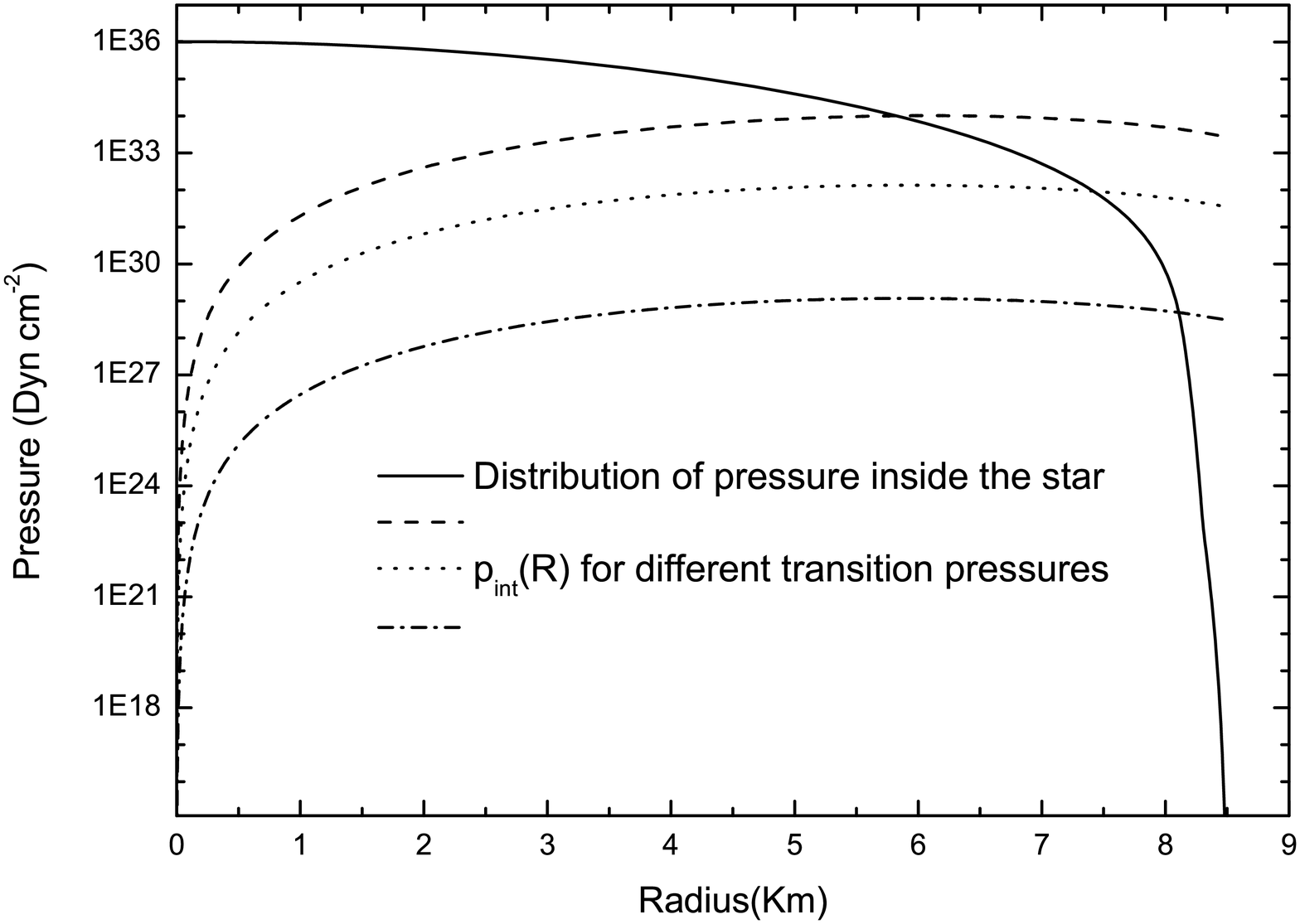}
\caption{The intersection between $p_{int}$, for different transition pressures, and p(R) determines the radius of the star}
\label{fig_1}
\end{figure}

\textbf{Order zero:} we obtain important conditions for the mass function and for the pressure of transition between core and \textit{crust}
\begin{subeqnarray}
M_{ext}=M_{int}+4\pi a^{2}\sqrt{1-\frac{2M_{int}}{a}}\varepsilon -8\pi
^{2}a^{2}\varepsilon ^{2}  
\slabel{59}\\
\frac{M_{ext}}{a^{2}\sqrt{1-\frac{2M_{ext}}{a}}}-\frac{M_{int}+4\pi
a^{3}p_{int}}{a^{2}\sqrt{1-\frac{2M_{int}}{a}}} = 4\pi \varepsilon 
\slabel{60}
\end{subeqnarray}
The equation (\ref{59}) can be interpreted as giving the total mass of the star to zero order ($M_{ext}$), as the addition of three terms, the first one representing the interior mass ($M_{int}\equiv 4 \pi \int_0^a \rho R^2 d R$), the second one representing the mass of the \textit{crust} and the third one being  a negative bounding energy term.

The equation (\ref{60}) shows that the introduction of a surface layer of energy density causes a discontinuity of the pressure at the border of the star. In the inner part of the surface the pressure is $p_{int}$ and outside the star the pressure is null.  In our model, $p_{int}$ represents the core-\textit{crust} transition pressure. We interpret (\ref{60})  as an equation giving the core-\textit{crust} transition pressure in terms of the total mass of the star, the interior mass, and $\varepsilon$.

This condition is used to obtain the  radius of the star, as can be seen in fig. \ref{fig_1}: from (\ref{60}) we define several functions $p_{int}(R)$, for different  transition pressures, and we study the intersection between these functions and the distribution of pressures in the interior of the star.

\textbf{First order: } we obtain continuity of the rate of rotation of the inertial frames, and an expression for the total angular momentum:
\begin{subeqnarray}
&&\omega(a)_{ext}=\omega (a)_{int}\equiv \omega (a)  
\slabel{conw}
\\&&J=\frac{a^{4}}{6}e^{\lambda(a)_{ext}}\left(e^{-\lambda(a)_{int}}\left[ \partial _{R}\overline{\omega }(a)\right]
_{int}+16\pi \varepsilon \overline{\omega_{c} }\right)
\slabel{conJ}
\end{subeqnarray}
where $\overline{\omega_{c} }=\Omega _{c}-\omega(a)$. 

It can be seen \cite{Hartle} that the first term of expression (\ref{conJ}) has the same sign as $\Omega$ (in our case always positive). However, the second term depends on the sign of $\Omega_{c}$. Hence, contra-rotating configuration are possible (the core and the \textit{crust} contra-rotate), and even,  we could find certain critical configuration in which the total angular momentum is null. 

\textbf{Second order: } we obtain the following conditions. 
\begin{equation}
\Delta\lbrack r^{\ast }(a,\theta )]=0
\end{equation}
where $r^{\ast } =r+\xi_0+(\xi_2+r \  k_2) P_2(\theta)$ is defined as in \cite{Hartle}, so both the mean radius $ \bar{r}^{\ast}=r +\xi_0$ and the eccentricity $e=[-3(k_2+\xi_2/r)]^{1/2}$ are continuous. We also obtain the next conditions
\begin{subeqnarray}
\nonumber
&\Delta&\lbrack e^{\lambda (a)}m_{0}(a)]=4\pi a^{2}\delta
\varepsilon _{0} + \frac{8\pi }{3}a^{4}e^{-2\psi (a)}\varepsilon \overline{\omega_{c} }^{2}+\\
&-&\frac{1}{2}\xi _{0}(a)\Delta \lbrack 3e^{-\lambda (a)}+e^{\lambda (a)}(8\pi a^{2}\rho (a)-1)]  
\slabel{con3}\\
\nonumber
&\Delta&\lbrack \frac{e^{\lambda (a)}}{F_{1}(a)}k_{2}(a)]-\frac{1}{%
12}a^{2}\overline{\omega }^{2}e^{-2\psi (a)}\Delta \lbrack \frac{F_{2}(a)}{%
F_{1}(a)}]\\ \nonumber &+&\frac{1}{24}a^{4}e^{-2\psi (a)}\Delta \lbrack \frac{e^{-\lambda
(a)}}{F_{1}(a)}\left( \partial _{R}\omega (a)\right) ^{2}]=\\&+&\frac{2\pi }{3}%
e^{-2\psi (a)}a^{3}\varepsilon \overline{\omega_{c} }^{2}-\pi a\delta
\varepsilon _{2}  \slabel{con4}\\
&\Delta&\lbrack e^{\lambda (a)}\xi _{2}(a)]=-\frac{8\pi }{3}e^{-2\psi
(a)}a^{4}\varepsilon \overline{\omega_{c} }^{2}  \slabel{con5}\\
\nonumber
&\Delta&\lbrack e^{\lambda (a)}F_{1}(a)m_{0}(a)]-8a\Delta \lbrack \frac{%
e^{\lambda (a)}}{F_{1}(a)}p_{2}^{\ast }(a)]\\ \nonumber &-&\frac{a^{5}}{6}e^{-2\psi
(a)}\Delta \lbrack e^{-\lambda (a)}\left( \partial _{R}\omega (a)\right)
^{2}]\\&+&\frac{1}{2}\xi _{0}(a)\Delta \lbrack F_{3}(a)]=-\frac{16\pi }{3}%
a^{4}e^{-2\psi (a)}\varepsilon \overline{\omega_{c} }^{2}  \label{con6}
\end{subeqnarray}
where we have defined $\Delta \lbrack f(a)]=f|_{\Sigma_{ext}}-f|_{\Sigma _{int}}$ and
\begin{subeqnarray}
F_{1}(a)&=&e^{2\lambda (a)}(1+8\pi a^{2}p(a))-1\\
F_{2}(a)&=&e^{\lambda (a)}(7+8\pi a^{2}p(a))-3e^{-\lambda (a)}\\
\nonumber 
F_{3}(a)&=&64\pi ^{2}a^{4}e^{3\lambda (a)}\rho (a)p(a)+8\pi a^{2}e^{3\lambda
(a)}(\rho (a)-p(a))\\&+&8\pi a^{2}e^{\lambda (a)}(3p(a)-\rho (a))-e^{3\lambda(a)}-3e^{-\lambda (a)}
\end{subeqnarray}
Basically, this conditions fix the jump in the second order perturbation of the mass function, and also fix the second order perturbation on the surface density ($\delta\varepsilon _{0}$ and $\delta\varepsilon _{2}$). Finally the last of the second order conditions fixes the value of the angular velocity of the \textit{crust}:
\begin{equation}
\Omega _{c}=\omega(a)  \pm (\Omega- \omega(a))  \label{con7}
\end{equation}
We obtain two possible configurations for the same star core and transition pressure: one with the \textit{crust} co-rotating with the interior fluid and with identical velocity, and a special configuration with contra-rotating \textit{crust}.  Current research on this strange contra-rotating configurations is under development, and it will be presented elsewhere. In what follows in this paper  we will use the co-rotating configuration.
 
\begin{figure}
\includegraphics[width=0.35\textwidth]{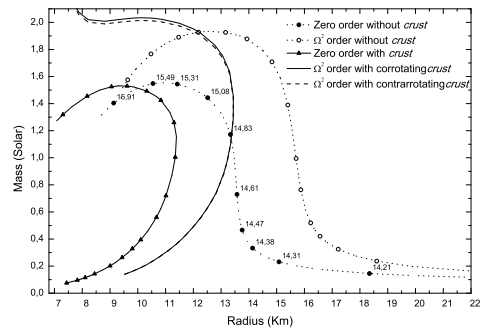}
\caption{Mass and radius of realistic neutron stars.}
\label{fig_2}
\end{figure}

In a summary, to obtain a model of a neutron star with a surface layer \textit{crust} in permanent slow rotation, for a given equation of state of the core, we have to fix three parameters. We choose the total mass of the star, the central density, and the core-\textit{crust} transition pressure. Then, we integrate the Einstein equations inside the star (using a realistic equation of state for the matter) and  we match this solution with the exterior solution found in \cite{Hartle}  using the junction conditions explained above. In fig. \ref{fig_2}, we show multiple configurations for both static stars and rotating stars, with a \textit{crust} of approximately 10$\%$ of the total mass. The angular velocity displayed is $\Omega^{2}_{S}=\frac{M_{ext}}{a^2}$ (mass shedding angular velocity \cite{Hartle}). On the static configurations without \textit{crust} we write the value $\log10(\rho_{c})$. We have used an equation of state for high pressures due to Glendenning  \cite{Glendening:2000}  (K=240 MeV, $\frac{m^{*}}{m}=0.78$ and $x_{\sigma}=0.6$) and the BPS equation of state for the lower pressures. Similar results have been obtained for the analitical fit of the Sly equation of state presented in \cite{Haensel:2004nu}.


\begin{center}
\textbf{\textit{Post-glitch} epoch of a Vela-type pulsar}
\end{center}

\begin{figure}
\includegraphics[width=0.32\textwidth]{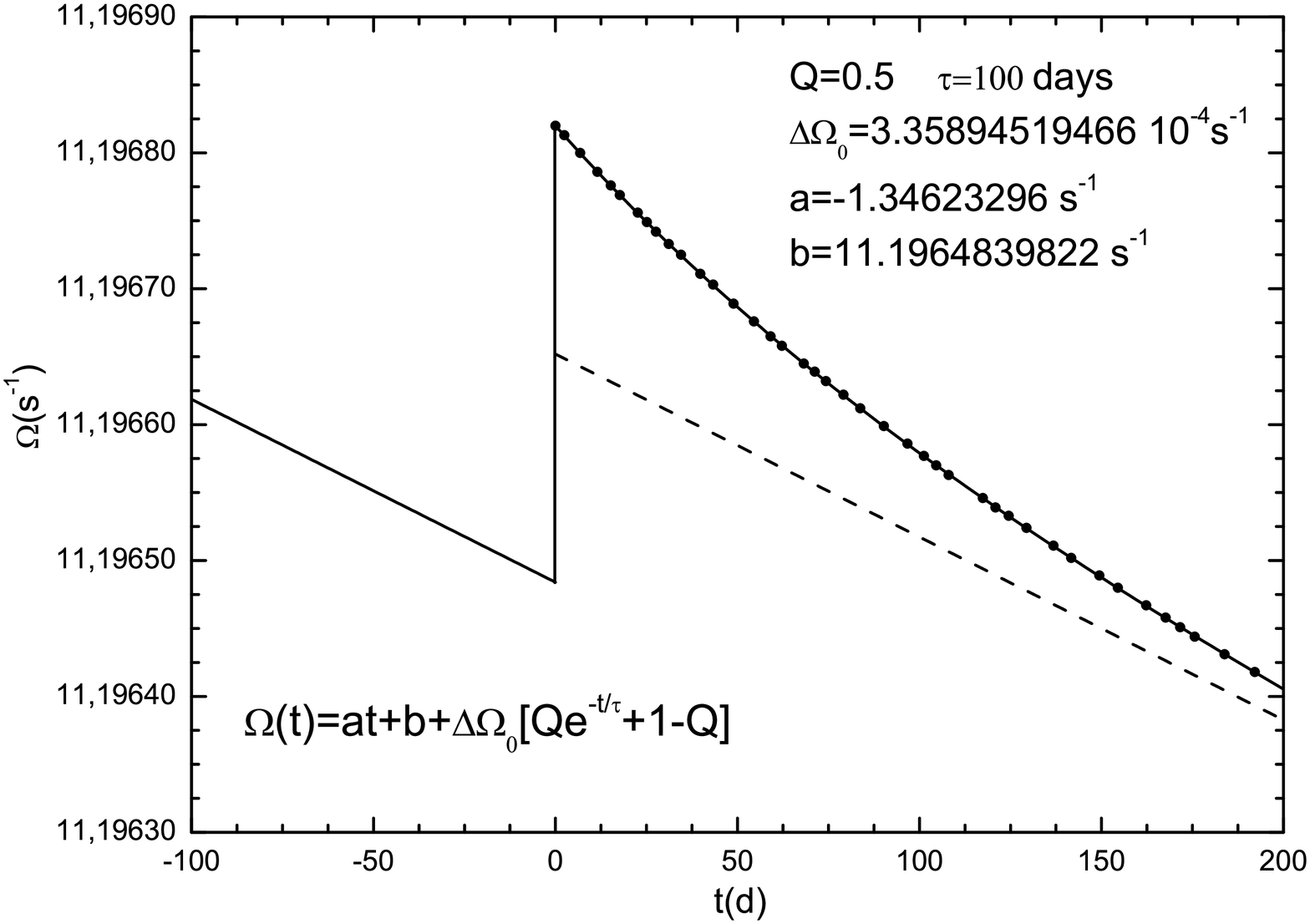}
\caption{Evolution of $\Omega$ on a typical Vela pulsar \textit{glitch}. The points correspond to the sequence of models.}
\label{fig_3}
\end{figure}

Now we are going to study the \textit{post-glitch} epoch of a Vela-type pulsar. The typical evolution of the angular velocity for a Vela-type pulsar during \textit{glitch} time and \textit{post-glitch} epoch is presented in fig. \ref{fig_3}  \cite{Shapiro}. It is interesting to note that in this problem there are two different scales of time, one related with the rotation period of the pulsar, and the other one  with the duration of the \textit{post-glitch} epoch. This fact allow us to describe the evolution of the pulsar in the \textit{post-glitch} epoch using a sequence of stationary and axisymmetric configurations with the appropriate parameters to follow the evolution of the angular velocity of the pulsar.
We assume that during the \textit{post-glitch} epoch neither the total mass nor, the central density of the star are changed. Hence, in our description,  the changes in the pulsar during this epoch are due to readjustment of the core-\textit{crust} transition pressure.

In order to control the core-\textit{crust} transition pressure, we assume that the \textit{crust} is thin enough to be treated as a surface layer \textit{crust}. Then to obtain the sequence of stationary and axisymmetric configurations mentioned above, we use our model of slowly rotating neutron star with superficial \textit{crust}. 

Let us describe our analysis. We assume that the neutron star has $1,44 M_{\odot}$ and central density $1.279\cdot10^{15} gcm^{-3}.$
At the beginning of the \textit{post-glitch} epoch this star has a
\textit{crust} with the 10$\%$ of the total mass. Since then, we adjust the
transition pressure to fit the angular velocity of the model to the evolution
shown in fig. \ref{fig_3}. The points in this figure correspond to the  elements of the sequence of models used to describe the evolution of the pulsar in the \textit{post-glitch} epoch.

\begin{figure}
\includegraphics[width=0.35\textwidth]{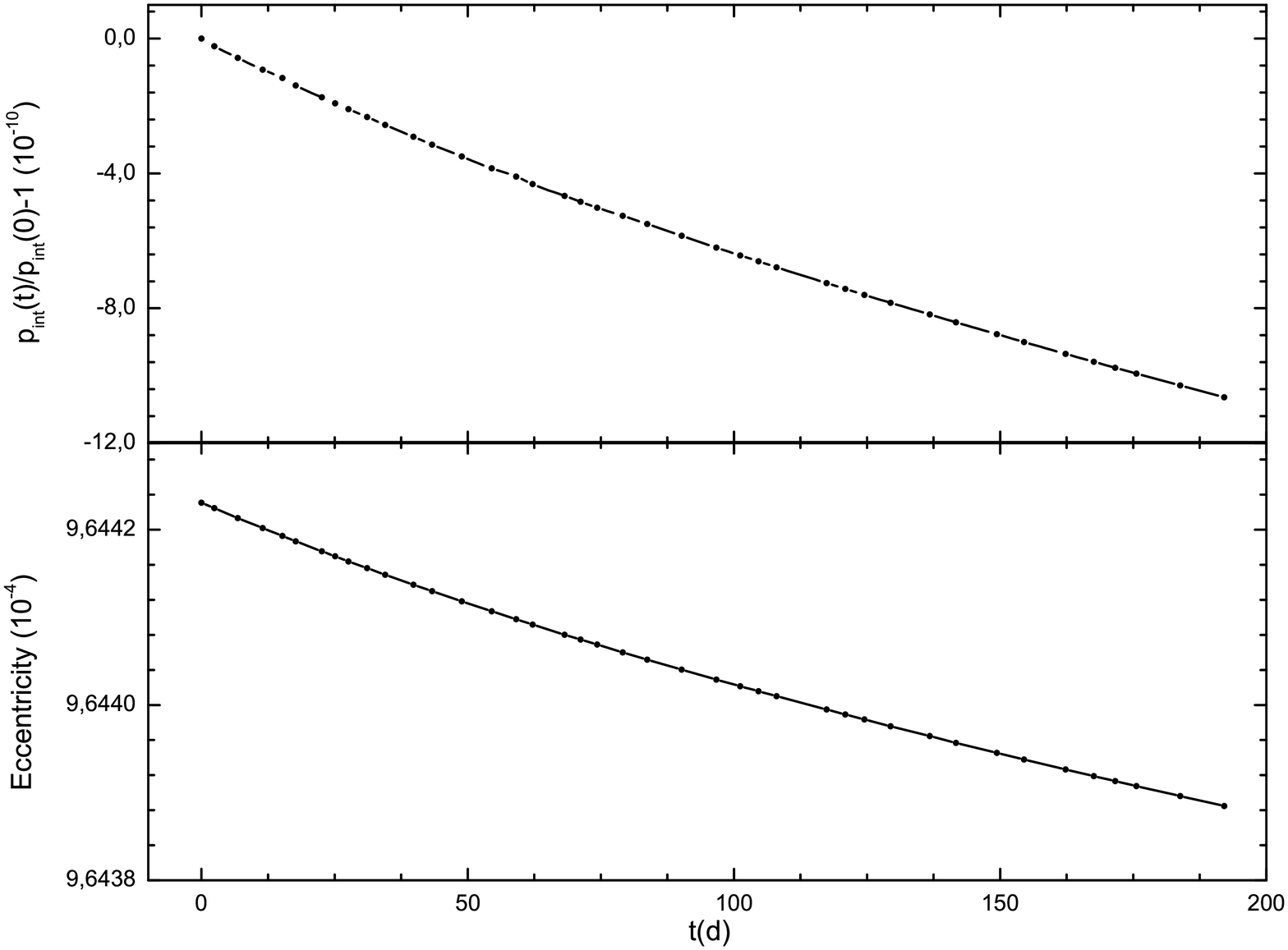}
\caption{Evolution of the core-\textit{crust} transition pressure and eccentricity during the \textit{post-glitch} epoch.}
\label{fig_4}
\end{figure}
\begin{figure}
\includegraphics[width=0.35\textwidth]{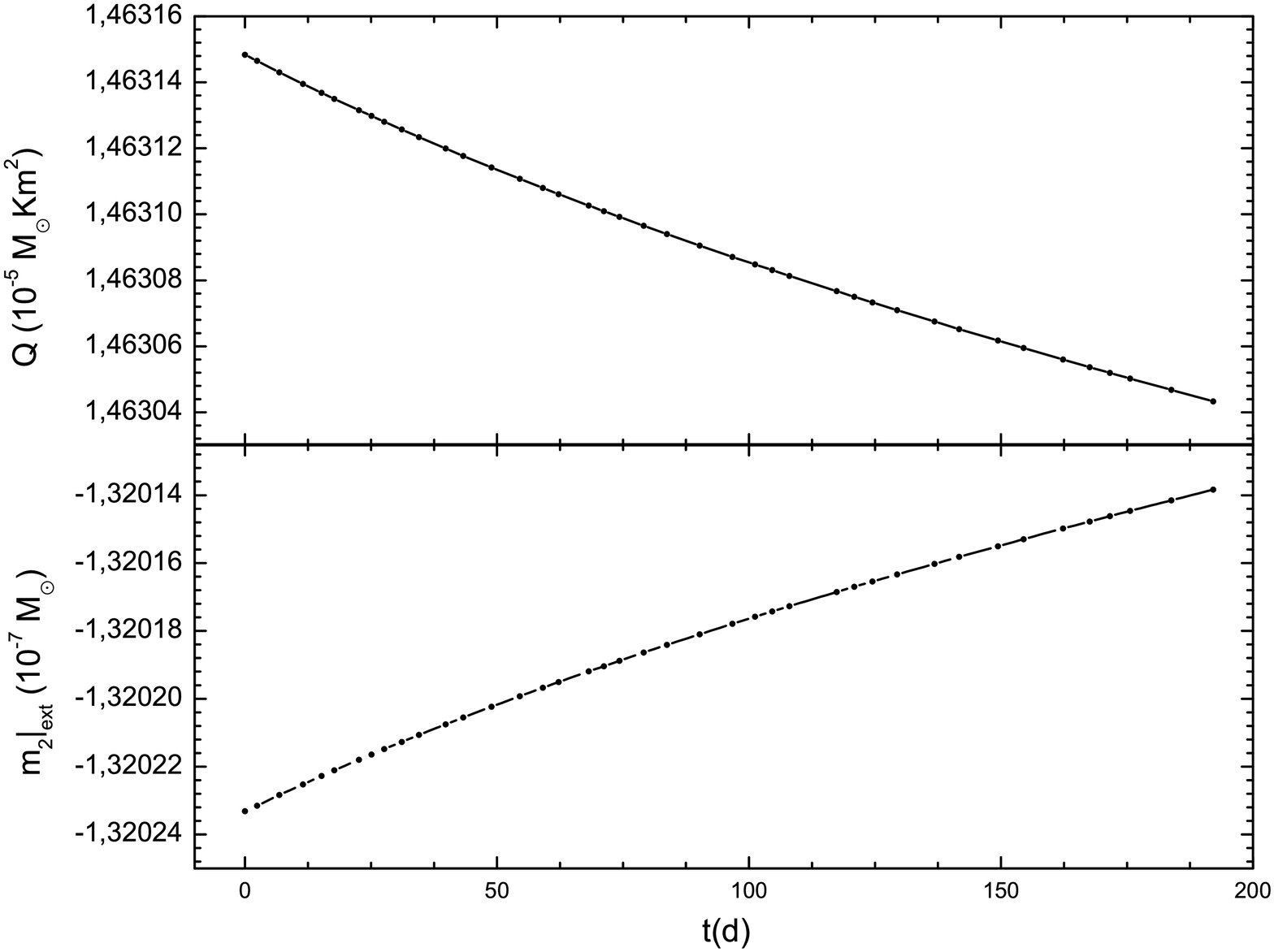}
\caption{Evolution of the surface matter distribution and quadrupolar moment during the \textit{post-glitch} epoch.}
\label{fig_5}
\end{figure}

We present our results in fig. \ref{fig_4},  that shows the evolution of the core-\textit{crust} transition pressure and eccentricity, and fig. \ref{fig_5} that shows the evolution of the surface matter distribution and quadrupolar moment. We obtain that   the 
core-\textit{crust} transition pressure decreases with time and causes a progressive reduction in the eccentricity, and a movement of matter on the surface of the star, from the equator  to the poles. Also,  the quadrupolar moment decreases in time all along the \textit{post-glitch}.

In consequence, we think that the \textit{post-glitch} epoch could be understood as a readjustment epoch for the core-\textit{crust} transition pressure. Our conjecture is that the temperature excess that the inner layers of the \textit{crust} have just after a  \textit{glitch}, would gradually dissipate during \textit{post-glitch} epoch until the \textit{crust} reaches its equilibrium temperature. These changes of temperature produce the evolution of the core-\textit{crust} transition pressure and the rest of the properties of the star.

The present work has been supported by Spanish Ministry of Science 
Project FIS2009-10614. The authors wish to thank F. Navarro-L\'erida for valuable discussions.


\end{document}